# Quantification and Modeling of Broken Links Prevalence in Hyper Traffic Websites Homepages

Ronan Mouchoux
Cyberdefense Department
University of South Brittany
Vannes, France
ronan.mouchoux@univ-ubs.fr

Laurent Moulin
Spartan Conseil
Paris, France
lmoulin@spartan-conseil.fr

Nicolas Striebig
XRATOR
Nantes, France
nicolas@x-rator.com

*Abstract*— Broken links in websites external resources pose a serious threat to cybersecurity and the credibility of websites. They can be hijacked to eavesdrop user traffic or to inject malicious software. In this paper, we present the first result of an ongoing research. We focus on the prevalence of broken links in external resources on home pages of the most visited websites in the world. The analysis was conducted on the top 88 000 homepages extracted from the Majestic Million rankings. 35,2% of them have at least one broken link. We also identify the common causes of these broken links and highlight improper implementation of testing phases to prevent such errors. We provide a formal model for the distribution of external links. At the next research step, we are exploring the potential impact on privacy of broken links by analyzing inherited traffic of purchasable expired domains.

**Keywords**—*Cybersecurity, Cybercrime, Web Security, External Attack Surface, Broken link, External Resource, Web Development, Web Monitoring*

## I. Introduction

Each day, approximately 330 000 people pass through Times Square in New York, one of the most visited physical places in the world. In the meantime, for Cyberplaces such as Facebook, Wikipedia, YouTube or Twitter, this number of daily visitors is the bare minimum. Those hyper-traffic websites became in two decades key gathering places for humanity.

The World Wide Web is based on interconnected resources through hypertext links. These links are essential to navigate between pages, load content and provide content and ergonomic rendering. However, over time, some of these resources may become unavailable, leading to broken links. Broken links can result from programming errors, typos, and expired domains hijacked links, among other reasons. In the best case, a broken link can affect the user experience, while in the worst case, it can pose a significant threat to the cybersecurity and the privacy of anyone visiting the website.

Given the importance of hyper-traffic websites in our life, it is critical to quantify the prevalence of broken links and identify the most common types of broken links. This paper presents an analysis of top 88 000 home pages extracted from Majestic Millions, identifying a total of 6 325 915 links to external resources, of which 1 052 244 are broken (16,63%).

Despite the existing research on broken links, there is a lack of studies focusing on the prevalence of broken links on high-traffic websites and a lack of studies on the potential weaponization of those anomalies. It paves the way for future research on opportunities brought by broken links for criminal purpose.

This paper is the first of a work in progress research. It provides a comprehensive and quantitative analysis of broken links' prevalence on hyper-traffic websites, as well as a formal model. Future research are focusing on the effective exploitability and weaponization of broken links by malicious actors.

The findings of this study demonstrate one vast opportunity offered to cybercriminal and predatory organization, allowing to better understand them and tackle them [1].

Web developers are called to be more diligent when conducting their pre-production testing and choosing their third-party libraries. Offensive Security professionals should rely on those findings to include broken links in their assessment and monitoring of their premises and their third parties, and treat them as exploitable vulnerabilities and not informative findings. Threat Hunters and Threat Intelligence analyst must consider broken links as a potential drive-by download infection vector when they reconstruct the threat actor intrusion path. Cybersecurity Product Designer may use the formal model to include the detection of outliers and anomalies in their product.

## II. Background

### A. Broken Links and the Web Landscape

Broken link (or dead link) is a web phenomenon observed since the 1990s. At that time, 3% of Digital Libraries objects became broken after one year [2]. On average, all URL links combined, the average lifespan is two years [3].

The problem has been covered academically primarily for the field of academic research or scholarship. Indeed, a scientific publication that cites its sources through hypertext links could pose problems of accessibility, and consequently credibility, peer review and reproducibility of results [4].

### B. Broken Links and Cybersecurity

The cybersecurity issue of broken links is discussed in academic paper but as an accessory or in a circumstantial manner. Patel et al. performed a security comparison of popular Content Managing Systems (CMS) to find out the strength and weakness of each regarding SQL injection, XSS and other popular web vulnerability [5]. Broken links are mentioned but more like an informative finding rather than a vulnerability that can be exploited. Bugliesi et al. mentioned broken links as a network solution for a remote attacker to collect a website session cookie [6].

In the best case, a broken link will simply result in a bad user experience or hamper the credibility of the website [7]. But in the worst case, it will pose a threat to the cybersecurity

of anyone visiting the website. Examples of hypertext link hijacking attacks are the following:

- Library Hijacking: Links to libraries removed and replaced by malicious one by ownership transfer [8].
- Domain Hijacking: Links to expired domains that can be registered or purchased [9].
- Subdomain Hijacking: Links to subdomains that are no longer in use and are vulnerable to takeover [10].

If a malicious actor successfully hijacks a called external resources, he could inject scripts executed by the victim's browser and harm the sites reputation and the visitor of the webpage [11].

In addition to malicious resources injection, broken links are a serious threat to privacy. If a malicious actor successfully takeover a called external resources, he could collect the source IP of the user, the source website of the user (the referrer), the potential data that were awaited by the broken resources [12]. The threat actor could also change the nature of the data awaited by the requester for weaponization purpose. On hyper-traffic websites, it is a cheap way for a malicious actor to set up a mass surveillance operation.

Finally, if the origin website is prone to discrimination (political, health, oppressed minorities, religion, sexual orientation, humanitarian) it is a candy bar for actor involved in blackmail or discriminatory intelligence. And a serious threat for the liberties and the life of oppressed and monitored visitors.

*C. Review of Cyber Incidents Leveraging Broken Links*

One of the first famous cases of the non-renewal of the domain is the Microsoft's passport[.]com, at the end of 1999. The domain name was renewed by a vigilant user, who was reimbursed soon after by Microsoft [13]. The same story is repeated in October 2003 with the domain hotmail[.]co[.]uk. Other major organizations such as Foursquare in 2010 [14] or Google Argentina in April 2021 [15] have failed to manage their domain name renewal properly, fortunately without any harmful effect.

On the criminal side of domain hijacking, in 2020, previously expired "parked" domains, purchased to simply be put up for sale, were being used to spread the Emotet malware. According to the article, 1% of parked domains were being used for malicious purposes [16]. In 2021, researchers at security firm Sucuri uncovered an attack exploiting the unrenewed domain tidioelements[.]com of the defunct WordPress plugin "visual-website-editor". Hackers registered this domain name to inject malicious JavaScript code into it [17].

Using third party library is a common practice in software development [18]. So, library hijacking by cybercriminals, leveraging broken links of abandoned open-source library project, are increasingly popular. In 2018, the U.S.-based bitcoin payments processor BitPay reported the compromission of their Copay wallet. The root-cause analysis pointed at a third-party abandoned JavaScript library that was took over by unidentified cybercriminal that pushes a malicious version [19]. In 2021, several cases of library hijacking occurred exploiting the popular Node.JS repository, NPM [20]. In 2021 there was a 650% year-on-year growth in cyber-attacks leveraging vulnerable and abandoned Open Source Software's supply chain [21].

*D. Our contribution*

Despite the existing research, there is a lack of studies focusing on the prevalence of broken links on hyper-traffic websites and associated modeling. This paper aims to address this gap in the literature by quantifying the prevalence of broken links on the homepages of high-traffic websites, identifying the types of external resources that are most prone to breakage and provide a formal model.

III. METHODOLOGY

To quantify the prevalence of broken links in hyper-traffic websites, we conducted a quantitative analysis using a web crawler. The collection process involved the following steps:

1. Selection of websites: We selected the top 88,000 websites from the Majestic Million, a stable top list source that provides a list of the most visited and most backlinked websites [22][23]. This selection criteria ensures that our sample represents popular websites in terms of traffic and backlinks.

2. Crawler configuration: We configured the web crawler to visit the homepages of the selected websites, to follow links to external resources.

3. Data collection: We inspected HTTP headers responses for each request on external link. Here, a broken link display anything but HTTP code 200, 301, 302 or 304. Inspecting the HTTP header's "Content-Type" field we also extract the type of resources. This collection occurred during April 2022.

4. Quality control: To ensure the quality of the data, we manually checked a random selection of one hundred reported broken links to verify that they were indeed broken. We also take the opportunity to qualify the root-cause of the broken link.

5. Data analysis: Our analysis involved statistical techniques and visualization methods to identify trends and patterns in the types of external resources linked to and the encountered errors.

The data collected and analyzed provides valuable insights into the types of external resources that are linked to and the types of errors that are encountered. These insights can be used to inform the development of strategies for preventing and mitigating the impact of broken links on website users.

IV. RESULTS

Our analysis of top 88 000 home pages extracted from Majestic Million revealed that of the 6 325 915 links to external resources, Images and JavaScript codes represent just over 70% of the external resources.

TABLE I.  TOP FIVE EXTERNAL RESOURCES BY TYPE

| *Resource Type* | *Number* | *Percentage* |
|---|---|---|
| Image | 2 536 692 | 40,1% |
| Script | 1 923 078 | 30,4% |
| Stylesheet | 752 783 | 11,9% |

| Resource Type | Number | Percentage |
|---|---|---|
| Font | 371 963 | 5.9% |
| XMLHttpRequest (XHR) | 354 251 | 5.6% |

Fig. 1. Top Five External Resources By Type

Of the 88,000 home pages, 30 960 have at least one broken link, which represents 35,2% of the total. Of the 6 325 915 links to external resources, 1 052 244 are broken, representing 16,63% of all links.

TABLE II. TOP FIVE BROKEN EXTERNAL RESOURCES BY TYPE

| Broken Resource Type | Number | Percentage |
|---|---|---|
| XMLHttpRequest (XHR) | 323 039 | 30,7% |
| Image | 292 524 | 27,8% |
| Script | 170 464 | 16,2% |
| Stylesheet | 68 396 | 6,5% |
| Fetch | 54 717 | 5,2% |

Fig. 2. Top Five Broken External Resources by Type

During the quality control phase, we manually reviewed a sample of broken links. We qualitatively notice that the most common cause of errors is programming errors, which are often the result of badly placed dots, inverted letters, forgotten separators between the domain and the resource, or loading the wrong resource. Domain names not renewed by advertising agencies or web design agencies are also a common issue.

On average, we have determined that the home page of a site has 72 dependencies. Approximately 60% are internal and 40% external. The distribution observed for all external references follows a gamma distribution, a shape parameter of 2,52 and a scale parameter of 30. We notice that the beginning of the distribution does not correspond to a gamma distribution: our hypothesis is that this correspond to the outliers that are not representative of the population.

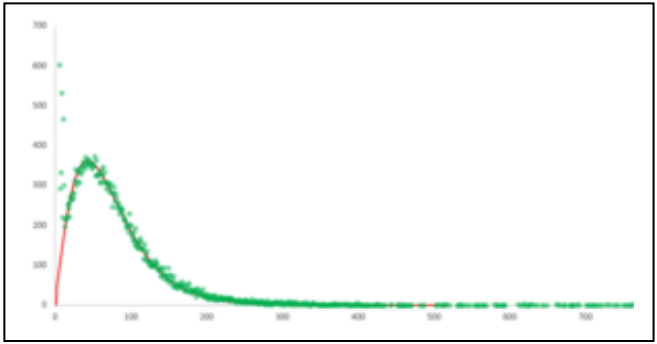

Fig. 3. Distribution of the number of all external references (y) on a homepage (x) in green, and in red a Gamma distribution (shape=2.52, scale=30).

The ability to model the distribution of external references per homepage provides the opportunity to detect easily and formally anomalies.

## V. DISCUSSION

Looking at our analysis, we can see that 35,2% of home pages have at least one broken link that may be taken over by a malicious person, regardless of the type of error. According to the company NetCraft, there were 1,13 billion websites in January 2021 [24]. Thus, about 398 million websites could be affected and present a risk. This number is likely a conservative estimate, as high-traffic websites homepages may have stricter security controls and scrutiny than medium or low-traffic websites.

One possible explanation for the presence of broken links is the lack of proper testing and quality assurance during the website development process. Our analysis showed that programming errors were a common cause of broken links. It suggests that developers are not conducting adequate testing prior to launching their websites. This highlights the need for more comprehensive testing and quality assurance practices to be implemented in the web development process.

We found that many broken links were the result of domain names not being renewed by advertising agencies or web design agencies, that were providing resources at the time of development. This could be due to poor follow-up of domain expiration dates or the cessation of the company's activities. This highlights the need for better management of its own domain names and to include critical third-party domain name in the security continuous monitoring.

Our study found that the average number of dependencies per website was 72, and that 40% of these were external resources. This means that external resources represent an additional risk vector for websites, as any issues with these resources could result in website downtime, poor User Experience, or cyber-attacks [11]. It is important for website owners to properly manage these external resources and ensure that they are necessary, reliable and secure.

XMLHttpRequest (XHR) and JavaScript code represented a little over 45% of the invalid external resources. This suggests that website owners and developers should pay particular attention to these types of resources when managing their websites, moreover when they are hosted on a third-party website. The ability for a threat actor to leverage a broken links requested scripts or XML is an invaluable intrusion and eavesdropping vector opportunity [25][26][27].

The total number of external references per homepages can be model using a gamma distribution. It offers opportunity to inform the development of statistical models and predictive algorithms.

There are several limitations to our study that must be highlight. First, our sample size was limited to 88 000 home pages, which may not be representative of the entire web. We yet choose on purpose to focus on the websites that are the most exposed to visitors, and thus have a greater criminal value in case of compromission.

Secondly, most hyper-traffic and high-traffic websites are now relying on Content Management System (CMS), such as Joomla, WordPress, and their associated plugin libraries. Thus, our analysis may not reflect the prevalence of broken links in hyper traffic websites, but the prevalence of broken links in their underlying technologies and methodologies used to develop them.

Additionally, our study focused only on the home page of each website, and did not analyze other pages that may impact the proportion we highlighted.

Finally, our study is limited by the fact that the web is constantly evolving, and the results may change over time.

VI. CONCLUSION

Our study focused on analyzing external resources of hyper traffic websites, in order to identify the presence of broken links and potential cybersecurity risks. Our analysis of top 88 000 home pages from Majestic Million revealed that 35.2% of the websites had at least one broken link, and 1,15% of all external resources were broken, which could be exploited by malicious individuals.

We also observed that images, XMLHttpRequest objects, and JavaScript codes represented a little over 70% of the total external resources. On average, the home page of a site had 72 dependencies, with 40% being external. We also noticed that the distribution of the number of dependencies per site followed a gamma distribution, which provide the opportunity to detected easily and formally anomalies.

Our research has several limitations, such as the sampling bias of our lists of domains and the temporal aspect of the web's volatility. Despite these limitations, our study provides insights into the cybersecurity risks of high traffic websites and the importance of testing phases to check programming errors and follow-up of domain expiration dates.

As a future direction, we plan to explore the analysis of purchasable expired domains present in broken links and make statistics on the inherited traffic. Our goal is to research to potential impact on privacy of broken links. We also intend repeating the test described in this study every year to analyze trends on external references and their associated anomalies.